\documentclass[iop]{emulateapj}
\usepackage{natbib,graphicx,amsmath,xspace}
\usepackage{color}

\newcommand{\mcg}{\mbox{MCG--5-23-16}\xspace}
\newcommand{\xmm}{\mbox{\emph{XMM-Newton}}\xspace}
\newcommand{\suzaku}{\mbox{\emph{Suzaku}}\xspace}
\newcommand{\nustar}{\mbox{\emph{NuSTAR}}\xspace}
\newcommand{\swift}{\mbox{\emph{Swift}}\xspace}

\begin{document}

\title{A long look at \mcg with \nustar: I- Relativistic Reflection and Coronal Properties}

\author{Abderahmen~Zoghbi$^{1}$, G. Matt$^{2}$, J.~M.~Miller$^{1}$, A.~M.~Lohfink$^{3}$, D.~J.~Walton$^{4}$, D.~R.~Ballantyne$^{5}$,  J. A. Garc\'{i}a$^{6,9}$, D.~Stern$^{4}$, M.~J.~Koss$^{7}$, D.~Farrah$^{8}$,
F.~A.~Harrison$^{9}$, S.~E.~Boggs$^{10}$, F.~E.~Christensen$^{11}$, W.~Craig$^{10}$, C.~J.~Hailey$^{12}$, W.~W.~Zhang$^{13}$}

\affil{$^{1}$Department of Astronomy, University of Michigan, Ann Arbor, MI 48109, USA}
\affil{$^{2}$Dipartimento di Matematica e Fisica, Universita degli Studi Roma Tre, via della Vasca Navale 84, I-00146 Roma, Italy}
\affil{$^{3}$Institute of Astronomy, University of Cambridge, Madingley Road,
Cambridge CB3 OHA, UK}
\affil{$^{4}$Jet Propulsion Laboratory, California Institute of Technology, Pasadena, CA 91109, USA}
\affil{$^{5}$Center for Relativistic Astrophysics, School of Physics, Georgia Institute of Technology, Atlanta, GA 30332, USA}
\affil{$^{6}$Harvard-Smithsonian Center for Astrophysics, 60 Garden Street, Cambridge, MA 02138, USA.}
\affil{$^{7}$Institute for Astronomy, Department of Physics, ETH Zurich, Wolfgang-Pauli-Strasse 27, CH-8093 Zurich, Switzerland}
\affil{$^{8}$Department of Physics, Virginia Tech, Blacksburg, VA 24061, USA}
\affil{$^{9}$Space Radiation Laboratory, California Institute of Technology, Pasadena, CA 91125, USA}
\affil{$^{10}$Space Science Laboratory, University of California, Berkeley, California 94720, USA}
\affil{$^{11}$DTU Space. National Space Institute, Technical University of Denmark, Elektrovej 327, 2800 Lyngby, Denmark}
\affil{$^{12}$Columbia Astrophysics Laboratory, Columbia University, New York, New York 10027, USA}
\affil{$^{13}$NASA Goddard Space Flight Center, Code 662, Greenbelt, MD 20771, USA}

\email{abzoghbi@umich.edu}

\begin{abstract}
\mcg was targeted in early 2015 with a half mega-seconds observing campaign using \nustar. Here we present the spectral analysis of these datasets along with an earlier observation and study the relativistic reflection and the primary coronal source. The data show strong reflection features in the form of both narrow and broad iron lines plus a Compton reflection hump. A cutoff energy is significantly detected in all exposures. The shape of the reflection spectrum does not change in the two years spanned by the observations, suggesting a stable geometry. A strong positive correlation is found between the cutoff energy and both the hard X-ray flux and spectral index. 
The measurements imply that the coronal plasma is not at the runaway electron-positron pair limit, and instead contains mostly electrons. The observed variability in the coronal properties is driven by a variable optical depth. A constant heating to cooling ratio is measured implying that there is a feedback mechanism in which a significant fraction of the photons cooling the corona are due to reprocessed hard X-rays.
\end{abstract}

\section{Introduction}
X-ray emission from AGN provides an excellent probe of the immediate vicinity of the central black hole.
The spectra of Seyfert galaxies at hard energies ($>10$ keV) are characterized by a powerlaw that rolls over exponentially at energies around $\sim100-200$ keV, along with a Compton reflection component. The powerlaw is the main X-ray source, produced by Comptonization of soft seed photons likely produced by the viscous dissipation of accretion energy \citep{1991ApJ...380L..51H}. Non-thermal Comptonization is not significant based on the detection of hard X-ray cutoffs and the non-detection of $\gamma$-rays \citep{1995ApJ...438L..63Z,2016A&ARv..24....2M}.

The best spectra prior to \nustar were provided by \emph{CGRO-OSSE} \citep{1993A&AS...97...21J}, \emph{BeppoSAX} \citep{1997A&AS..122..299B} and \emph{INTEGRAL} \citep{2003A&A...411L...1W}. Modeling of the spectra obtained using these missions \citep{1993ApJ...414L..93Z,2001ApJ...556..716P,2002A&A...389..802P,2013MNRAS.433.1687M,2014ApJ...782L..25M} provided some estimates of cutoff energies and reflection fractions, but offered only weak constraints on the physical models, mostly because the quality of the data was not high enough to break modeling degeneracies between the spectral index, the reflection strength and the cutoff energy. An additional difficulty was the requirement of simultaneous low energy ($<10$ keV) coverage that was not always available. \cite{2016MNRAS.458.2454L}, for instance, compare results from different published studies with \emph{BeppoSAX}, \emph{RXTE}/HXTE, \swift-BAT and \emph{INTEGRAL}, with lower ene,rgy coverage provided by several other instruments (mostly non-simultaneous), and found that estimates of the spectral index, the strength of reflection $R$ and the cutoff energy showed signi,ficant differences in values and correlations in the published work, even when analyzing the same datasets, mostly because of the low signal to noise ratio spectra above 30 keV.

The launch of \nustar \citep{2013ApJ...770..103H} offered a breakthrough for these studies. With its effective area and continuous energy coverage that includes the iron line at 6 keV and the Compton hump peaking at $\sim30$ keV, \nustar provides a better handle on obtaining accurate estimates (or stringent limits) of the Comptonization parameters. Many estimates have been published so far for individual Seyfert and radio galaxies including IC 4329A \citep{2014ApJ...788...61B}, Ark 120 \citep{2014MNRAS.439.3016M}, MCG-6-30-15 \citep{2014ApJ...787...83M}, SWIFT J2127.4+5654 \citep{2014MNRAS.440.2347M}, NGC~4945 \citep{2014ApJ...793...26P}, MCG--5-23-16 \citep{2015ApJ...800...62B}, NGC 2110 \citep{2015MNRAS.447..160M}, NGC~5506 \citep{2015MNRAS.447.3029M}, NGC~4151 \citep{2015ApJ...806..149K}, NGC~7213 \citep{2015MNRAS.452.3266U}, 3C~273 \citep{2015ApJ...812...14M}, NGC~5548 \citep{2015A&A...577A..38U}, 3C~390.3 \citep{2015ApJ...814...24L}, 3C~382 \citep{2014ApJ...794...62B} and MRK~335 \citep{2015MNRAS.454.4440W, 2016MNRAS.456.2722K}. Cutoff energies in the range $100-500$ keV are measured, and although these values are outside the energy band of \nustar (3--79 keV), the fact that the spectra starts rolling over well below the cutoff energy, and the effect of the cutoff on the reflection spectrum that cannot be mimicked by other parameters, both allow accurate measurements of the cutoff energy \citep{2015ApJ...808L..37G}.

There are many questions to address given these new observations. For instance, what is the geometry of the corona and what physical process controls the shape of the observed spectrum?. Population synthesis models for the X-ray background rule out significant emission above $\sim300$ keV \citep{2007A&A...463...79G,2014ApJ...786..104U}. Therefore, a cutoff should be observable in many sources. Additionally, depending on its compactness, the plasma cannot reach equilibrium at very high temperatures when photon-photon interactions become important. The production of electron-positron pairs acts as a thermostat, where increasing the power of the plasma produces more pairs rather than increases their temperature \citep{1984MNRAS.209..175S,1985ApJ...289..514Z}. Obtaining a measure of the electron temperature along with an estimate of the size of the corona is important in assessing to what extent these effects are important and help constrain the geometry of the corona.

Combining energy cutoff measurements from \nustar and size estimates from both spectral and timing information, \cite{2015MNRAS.451.4375F} found that the implied electron temperatures are close to the boundary of the region in the compactness-temperature diagram which is forbidden due to runaway pair production. This suggests that pairs are an important ingredient in AGN coronal plasmas. 

Theoretical models provide additional predictions that can be tested observationally. For example, in pair-free Compton-cooled coronae, an increase in cooling (keeping the power supplied to the electrons fixed) causes the temperature to drop, so $E_c$, the cutoff energy is anti-correlated with $\Gamma$, the photon index. On the other hand, in pair-dominated plasmas,
\cite{1994ApJ...429L..53G} found that $E_c$ is positively correlated with the observed photon index $\Gamma$ for electron temperatures $T_e<m_ec^2$ (where $m_e$ is the electron mass and $c$ is the speed of light), while the reverse trend is predicted above it. Even below this limit, $E_c$ can remain constant for different values of $\Gamma$\citep[e.g. Fig. 14b in][]{2002ApJ...578..357Z}.
These arguments apply to a single source, and should be observable in a sample of sources, as has been explored with \emph{BeppoSAX} and \emph{INTEGRAL} \citep[][]{2016MNRAS.458.2454L}, and is now being revisited with \nustar (Tortosa et al. \emph{in prep.}).

In this work, we use a long-look observation of \mcg ($z=0.0085$; $M=10^{7.9}M_{\odot}$, \citealt{2012A&A...542A..83P}) to attempt to address the question of cutoff variability and its relation to other parameters directly using a single bright object. We find that the cutoff energy is variable and shows interesting correlations with other parameters. The rest of this paper is organized as follows: Section \ref{sec:data_reduction} presents the data reduction and analysis, where we include data from \nustar, \swift \citep{2004ApJ...611.1005G} and \suzaku \citep{2007PASJ...59S...1M} to obtain a complete spectral picture between 1-79 keV. Section \ref{sec:spec_mod} presents detailed spectral modeling, first for the iron line band and then for the whole observed band. The implications of the results are discussed in section \ref{sec:discussion}. Results from the short time-scale variability will be published separately (Zoghbi et al. \emph{in prep.}).

\section{Observations \& Data Reduction}\label{sec:data_reduction}
\subsection{Observations Log}
Following the detection of both a cutoff energy \citep{2015ApJ...800...62B} and reverberation time delays \citep{2014ApJ...789...56Z} in the first observation of 2013, \mcg was observed again with \nustar for a net exposure of $528$ ks in early 2015. The new observation was split into three exposures, with the first two separated by 2 days, and the third taken 22 days later. The observation ID's, dates and exposures are shown in Table \ref{tab:obs_log}. Our analysis also includes data taken by \suzaku in 2013, simultaneously with the 2013 \nustar observation and also an earlier \suzaku observation taken in 2005 (see Table \ref{tab:obs_log}). Timing analysis of the 2013 \suzaku data was presented in \cite{2014ApJ...789...56Z}, while the spectral analysis is presented here for the first time.

\begin{table}
\begin{tabular}{|l|l|l|l|}
\hline
Satellite  & Obs. ID        & UT Date          & Exp. (ks) \\ \hline
\nustar    & 60001046002 (N2) & 2013-06-03    & 160       \\ 
           & 60001046004 (N4) & 2015-02-15    & 210       \\
           & 60001046006 (N6) & 2015-02-21    & 98        \\
           & 60001046008 (N8) & 2015-03-13    & 220       \\ \hline
\suzaku    & 700002010 (S0) & 2005-12-07    & 191         \\
		   & 708021010 (S1) & 2013-06-01    & 319         \\
		   & 708021020 (S2) & 2013-06-05    & 221         \\ \hline
\swift     & 0008042100[SW] & 2015-01-28    & $\sim 2$    \\ 
			&with SW = 2--11& to & each \\
		   &                & 2015-03-13    &             \\ \hline	   

\end{tabular}
\caption{A summary of the observations used in this work. }
\label{tab:obs_log}
\end{table}

In order to obtain spectral coverage below 3 keV simultaneous with the new \nustar observations, we requested snapshots with \swift XRT. A total of nine exposures were taken while \nustar was observing the source, in addition to one that was simultaneous with the first \nustar observation. The IDs of these observations are also shown in Table \ref{tab:obs_log}.

\subsection{Data Reduction}
\nustar data were reduced and analyzed using the \nustar data analysis software, which is part of {\sc heasoft v6.19}. The data were reduced by running the standard pipeline \texttt{nupipeline}. Source spectra were then extracted for modules A and B from circular regions 3 arcmin in radius centered on the source. Background spectra where extracted from source-free regions of the same size near the source. For the calibration files, we use {\sc caldb} release v20160502. In the spectral analysis, spectra from modules A and B are fitted simultaneously, allowing for a multiplicative constant between them.

The XIS spectra from \suzaku were reduced also using the relevant software in {\sc heasoft v6.19}. The initial reduction was done using \texttt{aepipeline}, using the {\sc caldb} calibration release v20160607. Source spectra were extracted using \texttt{xselect} from circular regions 3 arcmin in radius centered on the source. Background spectra were extracted from a source-free region of the same size, away from the calibration source. The response files were generated using \texttt{xisresp}. Spectra from XIS0 and XIS3 were checked to be consistent and then combined to form the front-illuminated (FI) spectra. Comparing front- and back-illuminated spectra by fitting an absorbed power-law region between 3 and 5 keV gives an index that varies by $\sim5\%$. Energies between 1.8 and 2.3 keV are ignored due to the calibration uncertainties associated with the CCD Si K edge. Data from other \suzaku instruments were not used because of low signal.

The \swift XRT observations were taken in the windowed timing mode except for SW2 and SW11 which were in imaging mode. The data were reduced using \texttt{xrtpipeline}. Source spectra were extracted from 
circular regions of 30 pixel radius (71 arcsec), and background spectra from similar regions away from the source. Observations SW2 and SW11 were taken in imaging mode and suffered some pileup. Therefore, the spectra were extracted from regions that excluded the central 3 pixels. We used the \swift\, {\sc caldb} release v20160121.

Spectral channels from \nustar, \suzaku XIS and \swift were grouped to have a minimum of one count per bin, and we use Poisson likelihood maximization in the modeling. Background spectra are handled using the W-statistics \citep{1996ASPC..101...17A}, where the background counts for each bin are considered fit parameters which can be solved for analytically as a function of other parameters. We use Poisson likelihood in order to be able to fit the swift low energy spectra and exploit the full resolution of \nustar at energies above 50 keV to constrain the high energy turnover. Using a Gaussian likelihood requires the channels to be grouped, which effectively removes some energy-dependent information at those energies.

\begin{figure}
\centering
\includegraphics[width=240pt,clip ]{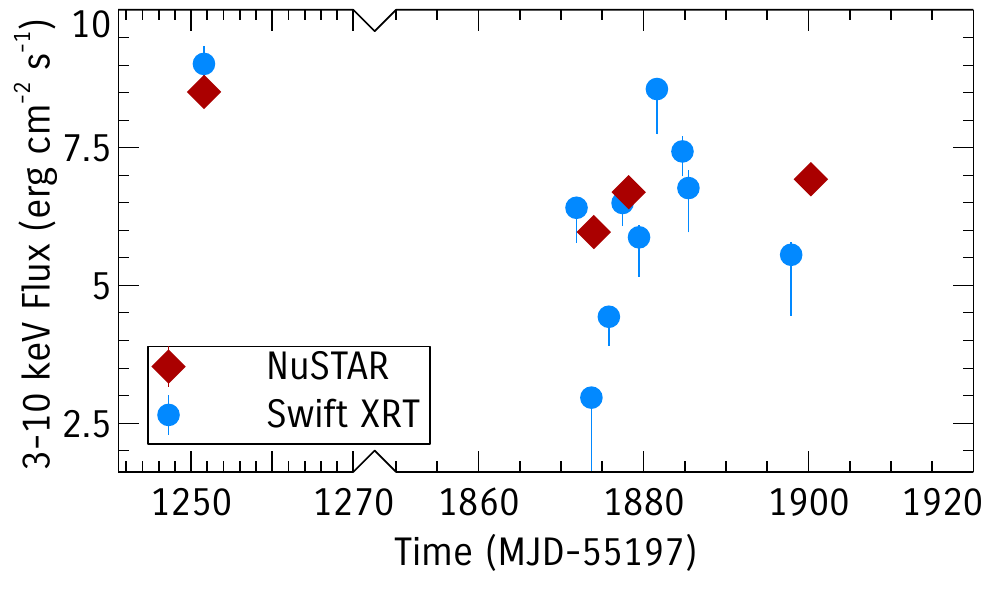} 
\caption{The long term light curve from the \swift monitoring (blue circles) along with the \nustar fluxes (red diamonds). Fluxes are obtained by fitting a powerlaw model to the 3-10 keV band. The S1-2 data from \suzaku have similar fluxes to the \swift point shown.}
\label{fig:long_lc}
\end{figure}

The long term light curve from the \swift monitoring is shown in Figure \ref{fig:long_lc}. The flux changes by a factor of $\sim3$ on a few days timescale, with the first 2013 (N2) observation having the highest flux. The \nustar observations sample the upper half of the flux variations observed with \swift. Given this variability and the simultaneity of observations, and for the purpose of constraining any variable column density at the host galaxy, we simultaneously fit observations that are taken within a day or so, or those having roughly the same 2--10 keV flux, modeling the \nustar and \suzaku data separately. Therefore, the following spectral groups are fitted with the same model, allowing for a multiplicative constant to account for cross calibration between instruments: [S0], [S1], [S2], [N2(A,B), SW8], [N4(A,B), SW3], [N6(A,B), SW6] and [N8(A,B), SW9]. We do not use SW2 as most of the counts are lost when correcting for pileup, removing its ability to constrain the column density below 3 keV. The cross calibration offset between the two \nustar modules is $<4\%$ is all cases, consistent with \cite{2015ApJS..220....8M}.

\section{Spectral Modeling}\label{sec:spec_mod}
All the spectral modeling is performed in {\sc xspec v.} 12.9.0n \citep{1996ASPC..101...17A}. The uncertainties quoted for the parameters are $1\sigma$ confidence, corresponding to $\Delta {\rm log}(Likelihood)=0.5$ (or $\Delta W=1$ where W is the W-statistic used in \textsc{xspec}), unless stated otherwise. The Galactic column in the direction of \mcg is $N_{\rm H} = 9\times10^{20}\, \rm{cm}^{-2}$ \citep{2005A&A...440..775K}, and it is included in all subsequent fits using the model \texttt{tbabs}. We start this section by showing a model-independent representation of the data, then focus on the iron line first by using phenomenological models to track the long term variability, then use a physical model in section \ref{sec:k_full}. We extend the analysis to the whole \nustar band in section \ref{sec:full_nu}.

\begin{figure}
\centering
\includegraphics[width=240pt,clip ]{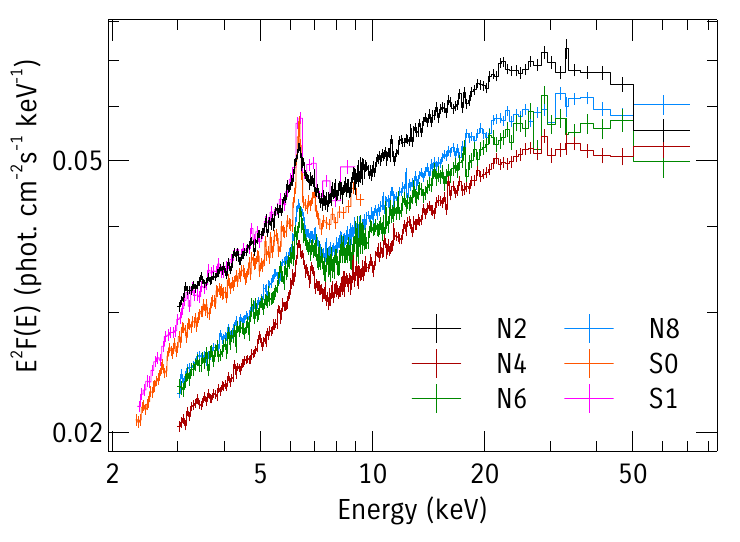} 
\caption{Unfolded $E^2F(E)$ plot for a representative subset of the spectra of \mcg from the \suzaku and \nustar campaigns. The nomenclature of the labels is defined in Table \ref{tab:obs_log}.}
\label{fig:nu_suz_eeu}
\end{figure}

Figure \ref{fig:nu_suz_eeu} shows the unfolded spectrum of \mcg. Data from \suzaku observations S0 and S1 (see Table \ref{tab:obs_log}), along with the four \nustar observations, plotted after factoring out the effective area of the detectors. The spectra have been re-binned to a minimum signal to noise ratio of 6 for display clarity. The spectrum of \mcg is characterized by strong iron K emission and a broad excess at 30 keV, characteristic of a reflection spectrum. \mcg is seen through a Compton-thin absorber ($N_{\rm H}\sim1.4\times10^{22}\, \rm{cm}^{-2}$), and little emission from the nuclear regions escapes below 1 keV. The observed spectrum at these energies (below 1 keV), as revealed by \xmm RGS spectra, is dominated by several emission lines superimposed on an unabsorbed scattered power-law continuum \citep{2007ApJ...670..978B}. They originate in a plasma in the Narrow Line Region. In the following \suzaku and \swift spectral modeling, we ignore spectral energies below 1 keV since they are not part of the nuclear emission. 

\subsection{The Fe Line Complex}\label{sec:iron_k}
The spectra of \mcg at the iron energies show both a narrow and a broad components \citep{2007PASJ...59S.301R}. The goal of this section is first to investigate whether the broad and narrow components are variable, and second to model the broad line with relativistic reflection models. All fits in this section are done in the 1--10 keV band.

\subsubsection{Long Term Variability}
We first attempt to check for column density variability using the \swift data alone. We fit an absorbed powerlaw to all the \swift spectra and track their changes. The column density showed some changes. However the strong degeneracy between column density and powerlaw index does not allow any firm conclusions to be drawn, and therefore analysis of all the spectral data is required.

\begin{figure}
\includegraphics[width=240pt,clip ]{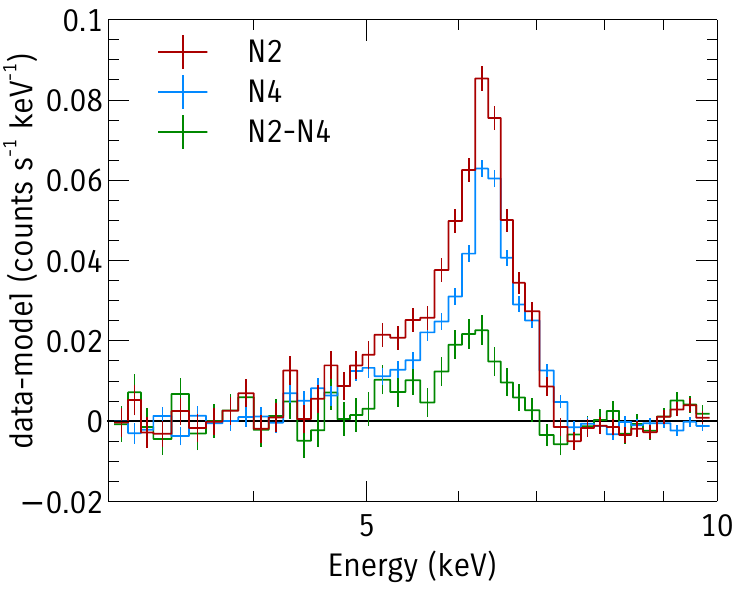}
\caption{The iron line from the N2 and N4 observations (the extremes in flux), with their difference. The spectra are plotted after subtracting a powerlaw model fitted to the 3--4 and 8--10 keV bands.}
\label{fig:diff_spec}
\end{figure}

To explore the changes in the iron line complex using a phenomenological approach, we show in Figure \ref{fig:diff_spec} the spectrum of the iron line from the highest and lowest flux \nustar observations (N2 and N4, respectively, which are separated by almost two years). The difference between the spectra is also shown. The spectrum in the iron line complex is clearly variable, and most of the variability is in the low energy wing of the line, not the core. The centroid energy for the N2 and N4 spectra are $6.30\pm0.01$ and $6.36\pm0.01$ keV, respectively, while the centroid from the difference spectrum is $6.17\pm0.05$ keV. \emph{It appears therefore that the strongest changes in the iron line are in the broad component and not the narrow component.}

\begin{figure}
\includegraphics[width=245pt,clip ]{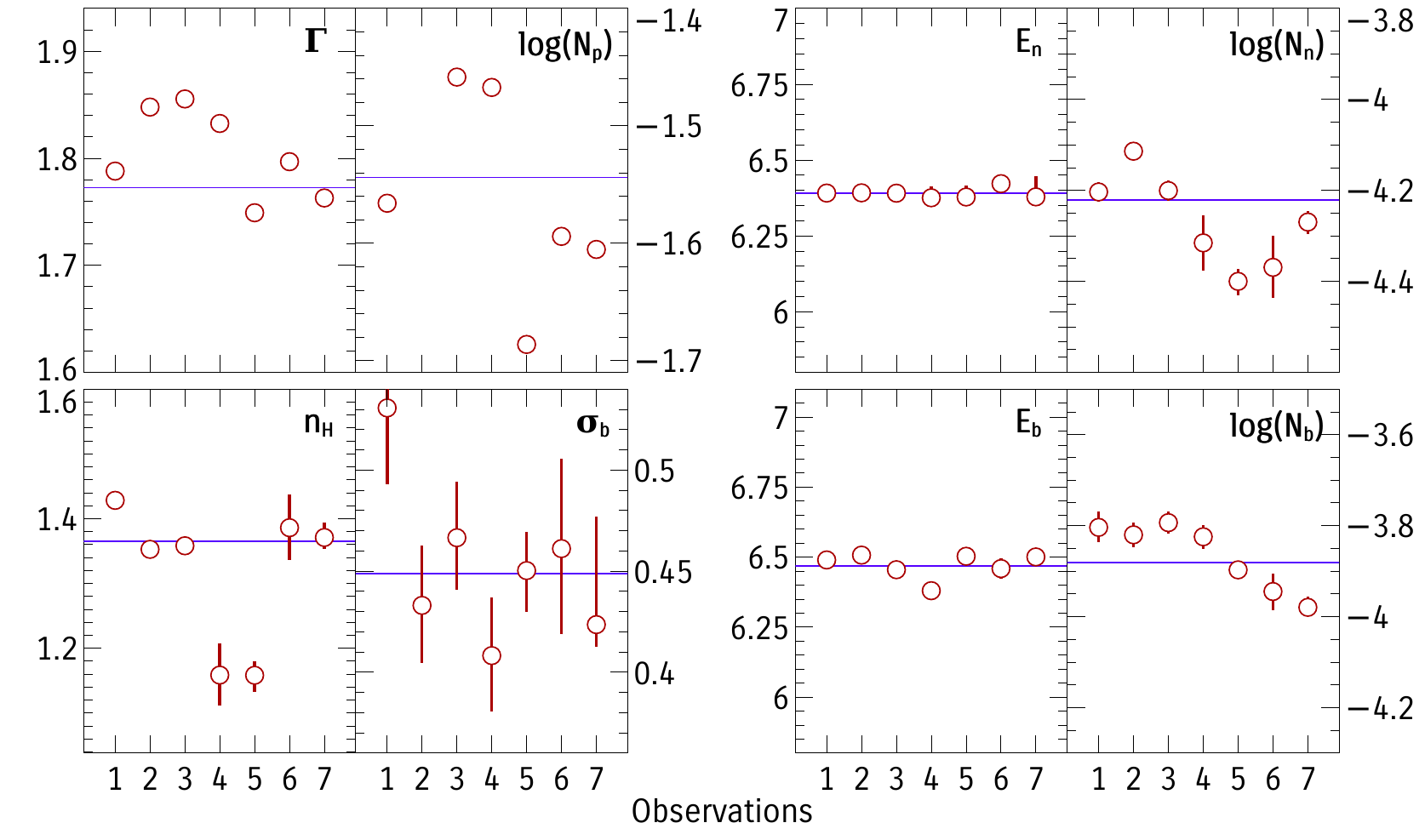}
\caption{Changes in the iron line complex inferred by using a powerlaw plus a narrow and broad Gaussian lines. Each tick in the x-axis is a single observation ordered as: S0, S1, S2, N2, N4, N6 and N8 respectively. The name of the parameter is shown in the top right of each panel. $\Gamma$ is the photon index. $N_p$ is the powerlaw normalization. $E_n$ and $E_b$ are the energies of the narrow and broad Gaussian lines respectively, and $N_n$ and $N_b$ are their normalizations. $\sigma_b$ is the width of the broad component.}
\label{fig:fit1b}
\end{figure}

To investigate this in a more systematic way, we fit the 1--10 keV band of the spectral groups using a powerlaw, a narrow and a broad Gaussian line. The width of the narrow line is fixed at the instrument resolution as neither \nustar nor \suzaku data are able to resolve it. Constraints to the column density for the \nustar data are provided by including \swift data, as discussed in section \ref{sec:data_reduction}. The resulting parameter changes are shown in Figure \ref{fig:fit1b}.

\nustar observation N2 was simultaneous with \suzaku S2, but there are clear systematic differences between them, both in the absolute flux values and in the photon indices. These are due to absolute calibration uncertainties between the two detectors and the uncertain cross calibration between the front- and back-illuminated detectors in \suzaku. The photon index differences are the main cause of the large difference in $N_{\rm H}$ between S2 and N2 (which are degenerate in the fits). Nonetheless, it is clear that most of the variability is in the powerlaw continuum flux and its photon index. The energy of the narrow component is consistent with a constant. In fact, a model where the line energy is fixed to the average is statistically preferred over a model where it is free to change.

We assess the significance of the parameter changes using the sample-corrected Akaike Information Criterion \citep[cAIC;][]{1974ITAC...19..716A,Burnham2011}, where we compare models in which the parameter of interest is either fixed or allowed to change between observations. For the narrow line, we find that fixing the line flux between observation gives a lower cAIC than allowing it to change, with $\Delta\rm{cAIC}=-2$ from each observation. This implies that the model with a fixed line flux is preferred. For the line energy, $\Delta\rm{cAIC}<0$ for the \suzaku-only data and $\Delta\rm{cAIC}>10$ when the \nustar data are included. In other words, the line appears to change \emph{only} between different instruments but is constant within the same instrument. This suggests that the line flux is physically constant and the observed changes are caused by instrument absolute calibration. For the column density, there appears to be significant changes between observations, even when using the same instrument ($\Delta\rm{cAIC}>15$ corresponding to a $>3.3\sigma$ significance). Changes in the parameters of the broad component are significant at the $3\sigma$ level when the narrow component is assumed constant. Degeneracies between the two reflection components are addressed in section \ref{sec:k_full} where we use full physical models.

\begin{table*}
\centering
\begin{tabular}{cccccccccccc}

\hline\hline

Obs & $N_{\rm H}$	& $\Gamma$	& $R$	& $\theta$	& $R_{\rm in}$	& $q$		&${\rm log}(\xi)$	&$N_{\rm r}$	&$N_{\rm x}$	& $E_c$\\ \hline\hline
\multicolumn{10}{c}{Fits to individual spectra between \bf{1--10} keV}\\ \hline

S0 	& 1.49(1)	& 1.84(3)	& 0.3(1)& 57(7)		& $<28$			& 2(1)	& 2.6(1)	& -3.25(1)	& -3.38(6)	& ...	\\

S1 	& 1.41(1)	& 1.89(2)	& 0.2(1)& 39(8)		& $<40$			& 3(1)	& 2.6(1)	& -3.10(1)	& -3.4(1)	& ...	\\

S2 	& 1.41(1)	& 1.90(1)	& 0.21(3)& 47(7)	& 77(20)		& $>6$	& 2.70(5)   & -3.19(1)	& -3.4(1)	& ...	\\

N2	& 1.24(5)	& 1.91(1)	& 0.29(8)& 34(7)	& 41(18)		& 5(2)	& 2.3(1)	& -3.18(1)	& -3.6(1)	& ...	\\

N4 	& 1.26(5)	& 1.83(2)	& 0.5(1)& 64(6)		& $<41$			& $<7$	& 2.5(9)	& -3.34(1)	& -3.5(1)	& ...	\\

N6	& 1.46(5)	& 1.88(2)	& 0.4(2)& 52(13)	& $<83$			& 2(1)	& 2.3(8)	& -3.29(1)	& -3.6(5)	& ...	\\

N8	& 1.34(5)	& 1.84(1)	& 0.29(7)& 45(8)	& $<33$		& 3(2)	& 2.3(8)	& -3.26(1)	& -3.5(1)	& ...	\\

\hline\hline
\multicolumn{10}{c}{Fits to individual spectra between \bf{1--79} keV}\\ \hline

N2	& 1.39(1)	& 1.87(1)	& 0.2(1)& 51		& $>50$			& 8(2)	& 2.5(5)	& -3.23(1)	& -3.45(1)	& 140(10)	\\

N4 	& 1.26(4)	& 1.77(1)	& 0.34(3)& ...		& $<68$			& 1(1)	& 2.7(1)	& -3.41(2)	& -3.84(5)	& 120(8)	\\

N6	& 1.41(5)	& 1.82(1)	& 0.25(7)& ...		& $<59$			& 1(2)	& 2.5(1)	& -3.33(1)	& -3.7(1)	& 146(18)	\\

N8	& 1.35(4)	& 1.77(1)	& 0.18(3)& ...		& $<73$			& 0.2(7)& 2.7(1)	& -3.33(1)	& -3.66(5)	& 124(7)	\\

\hline\hline
\multicolumn{10}{c}{Joint fit to all the spectra between \bf{1--79} keV}\\ \hline
   & $N_{\rm H}$& $\Gamma$	& $N_p$	& $\theta$	& $R_{\rm in}$	& $q$		&${\rm log}(\xi)$	&$N_{\rm r}$	&$N_{\rm x}$	& $E_c$\\ \hline\hline
N2	& 1.40(1)	& 1.85(1)	& -1.47(1)& 51		& 47(23)		& 1.6(6)& 2.66(3)	& -3.80(1)	& -3.60(2)	& 152(8)	\\

N4 	& 1.25(4)	& 1.78(1)	& -1.68(1)& ...		& ...			& ...	& ...		& -4.21(6)	& ...		& 107(7)	\\

N6	& 1.45(4)	& 1.83(1)	& -1.59(1)& ...		& ...			& ...	& ...		& -4.12(7)	& ...		& 149(14)	\\

N8	& 1.41(4)	& 1.79(1)	& -1.60(1)& ...		& ...			& ...	& ...		& -4.11(1)	& ...		& 130(6)	\\

\end{tabular}
\caption{Fit parameter for the \texttt{relxill+xillver} for the 1-10 keV fits for the \suzaku and \nustar datasets, and for 1-79 keV for the \nustar datasets. The 1$\sigma$ statistical uncertainty in the last significant figure is shown in brackets. $N$ is the normalization in units of ${\rm log}$(photons cm$^{-2}$ s$^{-1}$). Subscripts $r$, $p$ and $x$ are for \texttt{relxill}, \texttt{cutoffpl} and \texttt{xillver} respectively. $R$ is the reflection fraction measured as the ratio of direct to reflected fluxes between 20--40 keV. $R_{\rm in}$ is in units of gravitational radius $r_g$ and the inclination $\theta$ is in degrees. $q$ is the emissivity index, $\xi$ is the ionization parameter, $\Gamma$ is the photon index and $E_c$ is the cutoff energy. The observations in the first column are defined in Table \ref{tab:obs_log}.}
\label{tab:fits}
\end{table*}

\subsubsection{Relativistic Model}\label{sec:k_full}
The narrow Gaussian line is due to distant reflection from the Broad Line Region or the torus. To model it more physically, we replace the narrow Gaussian with the reflection model \texttt{xillver} \citep{2014ApJ...782...76G}. \texttt{xillver} includes self-consistently emission from Fe K$\beta$ and allows for the reflection to be ionized. Using the \texttt{xillver+powerlaw} model accounts for most of the residuals around 6.4 keV, including the Fe K$\beta$ line present in \mcg, also seen in previous observations \citep{2007PASJ...59S.301R}. The \texttt{xillver} model provides a better fit compared to \texttt{pexmon} \citep{2007MNRAS.382..194N}, for example, with $\Delta W\sim28$ per observation for the same number of degrees of freedom for \suzaku spectra where the K$\beta$ line is most clearly seen. This corresponds to $\Delta\rm{cAIC}=17$ and a significance of $>3.5\sigma$. The best fit ionization parameter of the reflector in this case is ${\rm log}(\xi)=0$, consistent with neutral reflection\footnote{$\xi$ in subsequent discussions is in units of erg s$^{-1}$ cm$^{-1}$}.

Continuing the analysis of the spectra in the 1--10 keV band, we next model the broad component of the iron line with a full relativistic reflection model. We use the \texttt{relxill} model (version v0.4a) \citep{2010MNRAS.409.1534D,2014ApJ...782...76G}. The reflection spectrum is a result of a hard X-ray source illuminating a constant density disk. The observer sees emission from both the illuminating source and the reflector. The reflection spectrum is convolved with a relativistic kernel to model the strong gravity effects of the black hole. As we will show, the inner radius of the disk is $>10 r_g$, and because the inner radius is degenerate with the black hole spin, we fix the latter at the maximum. The fact that the inner radius is $>10 r_g$, the exact fixed spin value has little effect on the fits. We use a single powerlaw emissivity profile, and assume that the directly observed component has the same spectral index as that illuminating the disk. The high energy cutoff is fixed at 300 keV, and it is modeled when fitting the whole \nustar band in section \ref{sec:full_nu}. The abundance of the inner and outer reflectors are assumed the same. We fit all seven spectra independently, and the results are summarized in Table~\ref{tab:fits}.

The reflection parameters are generally better constrained in the \suzaku spectra. This is because the resolution of \nustar in the iron K band does not allow the narrow and broad components of the line to be unambiguously separated. One effect of this is that the values of the iron abundance are not consistent between \suzaku and \nustar. The three \suzaku spectra suggested a value of about 1, while it is was around 0.5 (the model minimum) in the \nustar spectra. We therefore fixed the \nustar value at the \suzaku value. The uncertainties in Table \ref{tab:fits} are calculated from Monte Carlo Markov Chains (MCMC) as the $1\sigma$ standard deviation of the chains. We found when exploring the likelihood space that it is multi-modal, with multiple parameter combinations having close likelihood values near the maximum. MCMC is therefore well suited to explore this multi-modality. All the chains reported here were generated using the affine invariant ensemble sampler \citep{Goodman&Weare}. All the chains were run several times and long enough to ensure convergence. The convergence is assessed with both the autocorrelation of the chains and the stability of the chain variances.

\begin{figure}
\includegraphics[width=240pt,clip ]{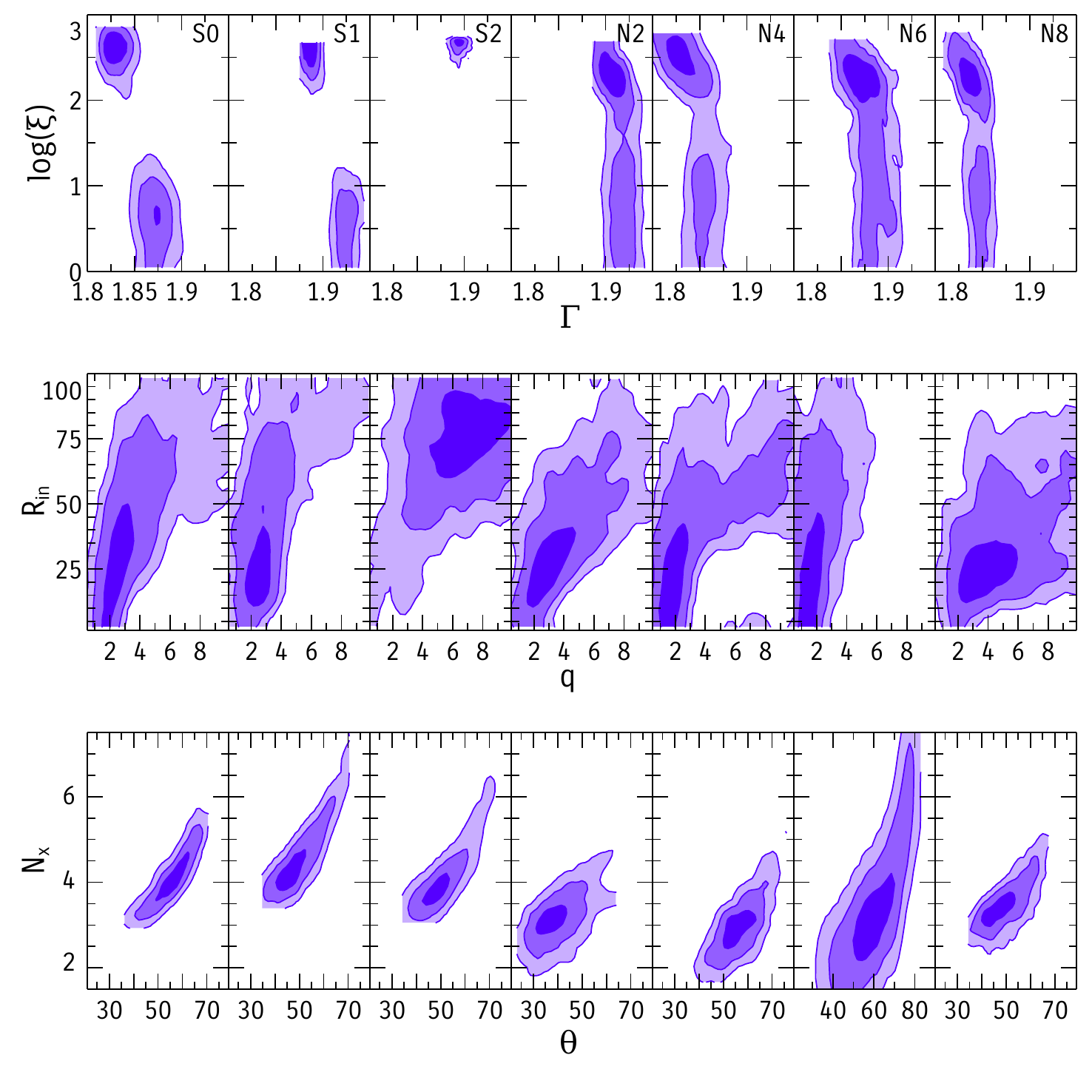}
\caption{Best fit confidence contours for the inner radius of the disk and emissivity profile of the disk as measured from \texttt{relxill} when fitted the spectrum below 10 keV. The top panel shows the confidence contours for the ionization parameters versus photon index, the middle panel is for the inner radius versus the emissivity index, and the bottom panel is for normalization of the \texttt{xillver} component versus the inclination angle.}
\label{fig:cfit2b_mcmc}
\end{figure}

Table \ref{tab:fits} quotes only the average values. There are however two general solutions in the modeling with different values for the ionization parameter $\xi$ of the relativistic reflection component (${\rm log}(\xi)\sim0$ and ${\rm log}(\xi)\sim2.7$). This is illustrated in the top panel of Figure \ref{fig:cfit2b_mcmc}, where we plot the confidence contours for the best fit parameters for (the log of) the ionization parameter versus the photon index for all the observations. Although the best fit value for the ionization parameter is ${\rm log}(\xi)\sim2.3-2.7$, lower values ($\leq1$) are also possible with a slightly higher value to $\Gamma$. Two parameters which are of interest are $R_{\rm in}$ and the emissivity index, which locate the emitting region relative to the central object. These two parameters are highly correlated in the modeling as shown by the middle panel of Figure \ref{fig:cfit2b_mcmc}. Although the best fit values are at $R_{\rm in}\sim20$ gravitational radii ($r_g=GM/c^2$) and $q\sim3$, lower and higher values for $R_{\rm in}$ are also supported by the data.

The confidence contours of the inclination $\theta$ and the normalization of the distant reflector $N_{\rm x}$ are shown in the bottom panel of Figure \ref{fig:cfit2b_mcmc}. The best fit inclination value in this case is $\theta\sim50^\circ$, but higher inclinations with stronger distant reflection are also supported by the data. The data in this case supports either a low inclination disk ($\theta<60^\circ$) with a relatively weak distant reflector or a highly inclined disk ($\theta\sim88^\circ$) and stronger reflection. Given that this object is seen through a Compton-thin rather than thick absorber, suggests an intermediate inclination, making the first solution more physically plausible. We note here that parameters reported in the analysis of the first \nustar observation \citep{2015ApJ...800...62B} are consistent with one of the local minima in the fit, with ${\rm log}(\xi)\sim0$. Our best fit has a value of ${\rm log}(\xi)\sim2.7$. Other parameters change accordingly, with reflection parameters not differing significantly (apart from the normalization). 

\begin{figure*}
\centering
\includegraphics[width=400pt,clip ]{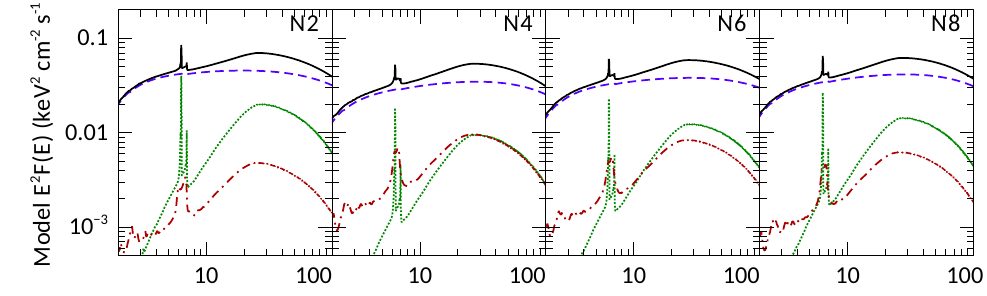}
\includegraphics[width=400pt,clip ]{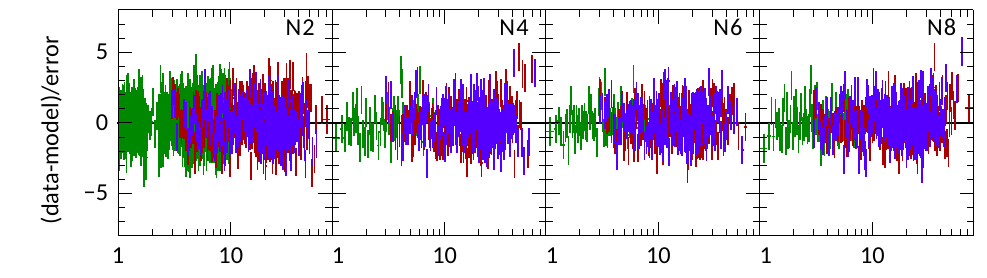}
\includegraphics[width=406pt,clip ]{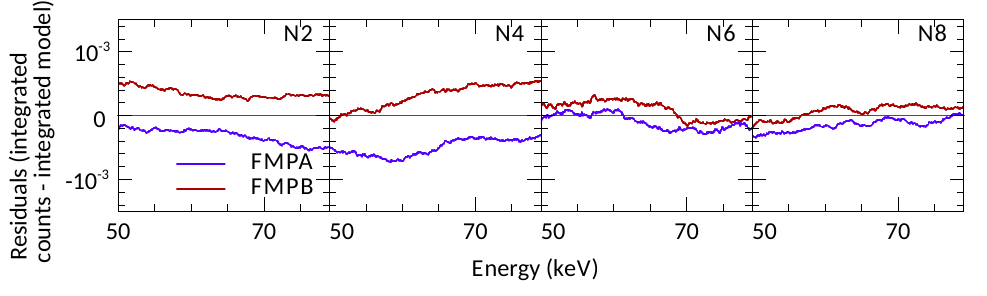}
\caption{\emph{Top}: The models fitted to the four \nustar epochs. The model consists of a powerlaw (blue, dashed), distant reflection (green, dotted) and relativistic reflection (red, dot-dashed), with their sum shown in solid black. \emph{Middle}: Fit residuals produced after rebinning the spectra. Red and blue colors are from the two \nustar modules, and the green is the low energy spectra from \suzaku (for N2) and \swift (for N4, N6 and N8). \emph{Bottom}: Residuals of integrated counts relative to the integrated model, a convenient way of visualizing the residual when the number of counts per bin is small.}
\label{fig:show_fit}
\end{figure*}

\subsection{Full Band Relativistic Model}\label{sec:full_nu}
Here, we extend the analysis to higher energies (up to 79 keV) and focus on the full \nustar data. Column density constraints are provided by \swift XRT for observations N4, N6 and N8, while for N2 we use the \suzaku observation S2 as it is of higher quality and it is simultaneous with N2. We model the spectrum with a model similar to that in section \ref{sec:k_full}, except that we fit for a powerlaw explicitly so we can track flux variations of individual components and allow for the cutoff energy to be a free parameter. The model has an \textsc{xspec} form: \texttt{tbabs*ztbabs* (relxill+cutoffpl+xillver)}. Allowing all parameters to be free showed the same strong correlations between the inclination and distant reflection flux discussed in section \ref{sec:k_full}, with the highly inclined disk being the best fit. Because this is physically unlikely, we fix the inclination at $\theta=51^{\circ}$, the weighted average from modeling the \suzaku spectra in section \ref{sec:k_full}. We found that the exact value for the inclination does not affect the following results significantly. The main effect is that a lower (higher) value causes the distant reflector to be weaker (stronger), as already suggested by the bottom panel of Figure \ref{fig:cfit2b_mcmc}.

The best fit parameters from modeling the whole \nustar band are shown in Table \ref{tab:fits}. The best fit model and residuals are shown in Figure \ref{fig:show_fit}. The residuals in the middle set of panels in Figure \ref{fig:show_fit} have been binned so that the signal to noise ratio of every spectral bin is at least 6. This is done so the residual plot is meaningful. The residuals at the high energy part of the spectra are shown in the bottom set of panels, plotted as the residuals of integrated (unbinned) data to the integrated model. This is a convenient way of plotting to account for the fact that bins at these energies have small number of counts. The bottom set of panels show that the deviations between the model and the data are comparable to the deviations between the two \nustar modules due to counting noise or cross calibration uncertainties.

The goodness-of-fit statistic is estimated using Monte Carlo simulations. We start from the best fit parameters and generate a large number of parameters drawn from the MCMC chains. For each parameter set, a spectrum is faked using \texttt{fakeit} in \textsc{xspec}, taking into account counting noise. The faked spectra are then refitted with the model and a distribution of fit statistics is produced from the resulting fits. The fraction of simulated data that have a fit statistic that is at least as good as the observed value are 0.96, 0.59, 0.21 and 0.4 for observations N2, N4, N6 and N8, respectively. These goodness parameters correspond to the probability of rejecting the null hypothesis corresponding to the best fit model. These fits are very good given the high quality data. The high value for the N2-S2 combination is due to cross calibration uncertainties between \nustar and \suzaku and between the front- and back-illuminated detectors in \suzaku.

Most of the parameters remained similar to those in section \ref{sec:iron_k} and Figure \ref{fig:cfit2b_mcmc}. 
$R_{\rm in}$, $q$ and the parameters of \texttt{xillver} were all consistent with being constant between observations. Therefore, and in order to obtain further constraints on the variable parameters, we fit all four \nustar observations (and the matching \suzaku and \swift datasets) together, allowing only parameters that showed variability in the individual fits to vary. These parameters are: $N_{\rm H}$, $\Gamma$, $E_c$ and the normalizations of \texttt{relxill} ($N_r$) and \texttt{cutoffpl} ($N_p$). The best fit parameters in this case are presented in Table \ref{tab:fits}.

\begin{figure*}
\centering
\includegraphics[width=400pt,clip ]{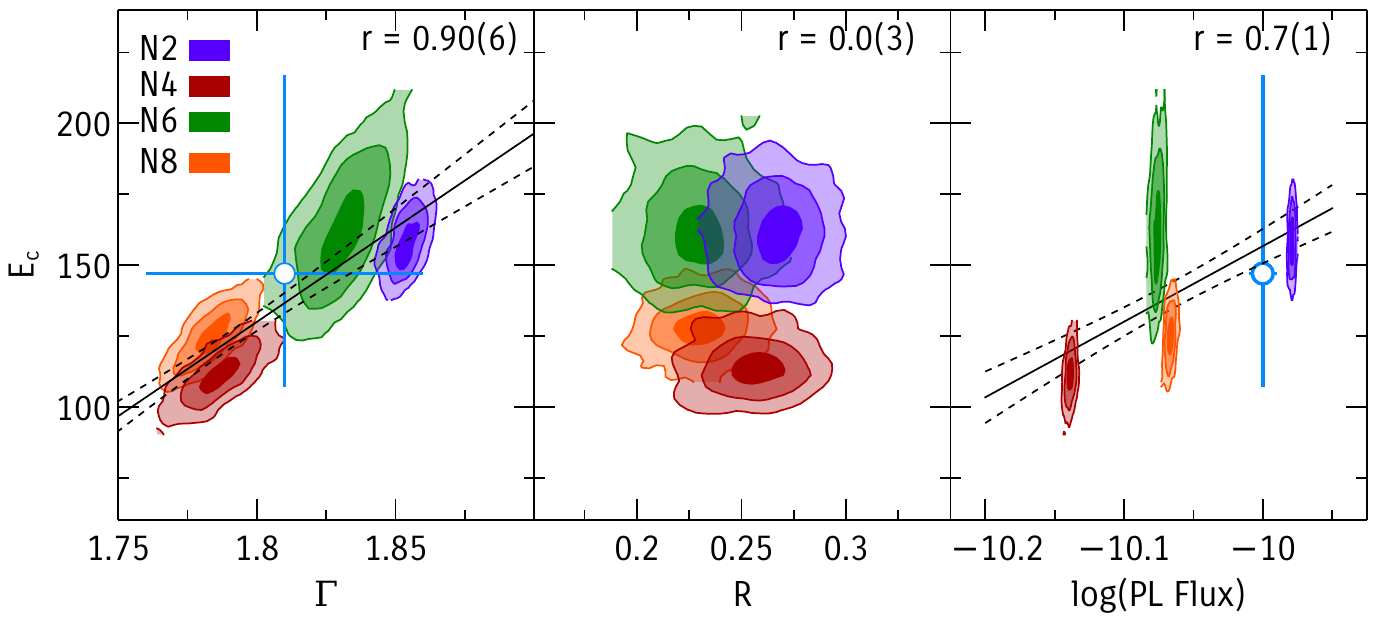} 
\caption{Changes in the parameters describing the primary X-ray continuum. The contours are the 1, 2 and 3 $\sigma$ confidence intervals for each parameter. $R$ is the reflection fraction. The solid line in the left and right-most panels is the best fit linear model and the dashed lines are the $1\sigma$ uncertainty in linear model. The Pearson correlation coefficient is quoted in the top-right corner of each panel. $E_c$ is in keV, the powerlaw flux $F_p$ is in units of ergs cm$^{-2}$ s$^{-1}$ and is measured between 2--10 keV. The blue points show a single measurement from a \emph{BeppoSAX} taken in 1998 \citep{2002A&A...389..802P}. A linear fit to the $E_c-\Gamma$ relation gives $E_c = (665\pm104) \Gamma +(-1067\pm186)$ and a linear fit to the $E_c-F_p$ relation gives $E_c = (267\pm58) F_p + (2822\pm586)$. The observation labeling N2-N8 is defined in Figure \ref{tab:obs_log}.}
\label{fig:ecut_var}
\end{figure*}

The column density appears to change between observations in a way that is not directly related to the observed flux. The variability is however only marginally significant. The 99.5\% confidence limits on $N_{\rm H}$ are consistent with a constant column. The remaining parameters change significantly between observations. The results of their variability is summarized in Figure \ref{fig:ecut_var}, which shows the best fit parameter confidence contours for $\Gamma$, reflection fraction\footnote{What we refer to as reflection fraction here is reflection strength in the nomenclature of \cite{2016A&A...590A..76D}.} $R$ and the 2--10 keV flux of the powerlaw component ($F_p$), plotted against the high energy cutoff.
Although flux from both the powerlaw and reflection components change, their ratio remains relatively constant. The difference between high- and low-flux model spectra in this case resembles the shape of the relativistic reflection (plus a powerlaw), and it directly explains the observed difference spectrum between N2 and N4 shown in Figure \ref{fig:diff_spec}. 

The photon index is correlated with the high energy cutoff $E_c$ within individual spectra. This is a consequence of the model parameterization when the data above $\sim50$ keV have a relatively low signal to noise ratio. Additionally, both $\Gamma$ and the continuum flux appear to be correlated with the cutoff energy when all four observations are considered. The reflection fraction is independent of cutoff energy. To quantify these correlations, we use the MCMC chains already calculated to calculate the Pearson correlation coefficient $r$. We find correlation coefficients of 0.90(6), 0.0(3) and 0.7(1) for the relations between $E_c$ and $\Gamma$, $R$ and $F_p$, respectively. The number in bracket is the uncertainty in the last significant digit taken as the $1\sigma$ sample standard deviation.

We emphasize that $R$ in Figure \ref{fig:ecut_var} is the reflection fraction from the \emph{relativistic} reflection. The reflection fraction from the distant reflector (not plotted), increases with $E_c$, driven by the $F_p-E_c$ correlation and the fact that the flux from the distant reflector is constant. Figure \ref{fig:ecut_var} also shows one measurement from a \emph{BeppoSAX} taken in 1998 \citep{2002A&A...389..802P}. That measurement appears to follow the same trends we observe, albeit with larger uncertainties.

We note here that the two, low and high $\xi$, solutions found when fitting data below 10 keV (section \ref{sec:iron_k}) no longer give a comparable goodness of fit. The higher $\xi$ solution gives a better fit by $\Delta W>30$, corresponding to a significance of $>3.7\sigma$ per observation. This solution is therefore preferred over that reported in \cite{2015ApJ...800...62B} for observation N2. Forcing the $\xi\sim0$ solution gives, in addition to a worse fit, lower cutoff energies but the correlations in Figure \ref{fig:ecut_var} hold.

We also note the low $N_{\rm H}$ value for N4. It is unlikely that the column density changes significantly in days time-scale. We therefore tested tying the $N_{\rm H}$ values between observations. We found a slightly worse fit ($\Delta W\sim14$ or a significance of $\sim 2\sigma$), with no large changes in $E_c$. The reason is that \nustar data quality is much better than \swift, so forcing a new $N_{\rm H}$ does not affect the \nustar data significantly but makes the \swift fit slightly worse.

\section{Discussion}\label{sec:discussion}

\subsection{X-ray reflection and the inner disk}
The most detailed analysis of the broad component of the iron line in \mcg prior to this work was presented in \cite{2007PASJ...59S.301R}. The results presented here regarding the reflection spectrum are consistent with that analysis. Taking the best fit parameters for the relativistic reflection suggest a truncated disk at $\sim40 r_g$ with a emissivity index close to a standard non-relativistic value of 3. There is, however, a strong degeneracy between the inner radius and emissivity index parameters, such that the data also supports (at the 99\% confidence level) a disk that extends down to $R_{\rm in}<6r_g$, with a flatter emissivity index ($q<2$). X-ray reverberation lags detected in this source, taken at face value suggest that the latter solution is preferred \citep{2014ApJ...789...56Z}.

The reflection component, producing the narrow core of the iron line and a considerable fraction of the reflection hump at 30 keV, appears to be constant across the two year period spanned by the data, unless the inclination of the disk changes, which is unlikely. This is not surprising if emission comes from material far from the central source, thereby smoothing out any variability. A consequence of this observation is that the reflection fraction from this component \emph{varies}, even when the flux and index of the illuminating source are the only parameters that change, \emph{not} the geometry of the system. A correlation of the distant reflection strength $R$ with the flux in a single object is therefore naturally explained \citep[e.g.][]{2002MNRAS.336.1209M}.

The flux from the inner reflector on the other hand, tracks closely variations of the illuminating source. The strength of the reflection (measured as the ratio of fluxes between 20 and 40 keV) remains constant, a result also seen in other objects \citep[e.g.][]{2016ApJ...821...11L}. Combining this with the fact that the column density changes very little, and also the lack of significant changes in the relativistic reflection parameter, gives a picture in which the flux changes seen in this source (e.g. Figure \ref{fig:long_lc}) are driven by \emph{intrinsic} flux fluctuations in the primary source, which are closely matched, on days to months timescales, by variations in the flux of the relativistic reflection component. This seems to be the long term extension of the relativistic reverberation signatures seen on short time-scales in this object \citep{2014ApJ...789...56Z}.

\subsection{Plasma properties}
Modeling the reflection spectrum properly allows us to extract information about the Comptonization process. We find that most of the variability between observations is due to changes in the \emph{flux} of the primary powerlaw component. Its photon index and cutoff energy are also found to be significantly variable, while the relativistic reflection component remains constant in shape and constant in flux \emph{relative} to the primary component. The primary continuum changes in a structured manner, as indicated by the high correlation coefficients between the photon index and flux with the cutoff energy. 

It is known that in the simple cutoff powerlaw model, the flux, index and cutoff energy can be correlated by construction. This is apparent in the elongated contours shown in Figure \ref{fig:ecut_var} from individual fits. It should be noted however, that the correlations \emph{between} observations are robust, in a sense that if the parameters from one observation are fixed at the best fit parameters of another observation, the fit significantly worsens. Another way to see this is to observe that in the $\Gamma-E_c$ and $F_p-E_c$ plots in Figure \ref{fig:ecut_var}, the elongated contours are \emph{not} parallel to the observed correlations, indicating that although the parameters might be correlated \emph{within} a single spectrum, their correlations \emph{between} observations are robust.

\subsubsection{$\Gamma-E_c$ Correlation}\label{sec:gam_ec}
Previous results on possible $\Gamma-E_c$ correlations in \emph{samples} of objects were not conclusive. \cite{1999NuPhS..69..481P} first noted a possible $\Gamma-E_c$ relation using two \emph{BeppoSAX} observations of NGC~4151. \cite{2001ApJ...556..716P} found a weak relation with six Seyfert Galaxies, and a relatively stronger relation was found by \cite{2002A&A...389..802P} using a slightly larger sample. Using \emph{INTEGRAL} data, \cite{2013MNRAS.433.1687M} reported a weak relation while \cite{2014ApJ...782L..25M} reported no relation. It appears therefore that as far as a sample of AGN is concerned, there is at most a weak relation between $\Gamma$ and $E_c$.

The data for individual objects is less clear. Mostly because of the difficulty of obtaining high quality data in single epoch observations, with low energy coverage. We note that from the \emph{INTEGRAL} study of \cite{2013MNRAS.433.1687M}, who analyzed separate observations of individual objects, in almost all cases of objects with multiple observations, flatter spectra are accompanied by small cutoff energies. The uncertainties in the parameters are, however, large. Using \nustar, \cite{2014ApJ...794...62B} analyzed two observations of the radio galaxy 3C~382 in two flux intervals. The low flux observation had a flatter spectrum and a higher cutoff energy compared to the higher flux observation (i.e. opposite the trend seen in \emph{INTEGRAL} data and seen here in \mcg). Also using \nustar data, \cite{2016MNRAS.456.2722K} found a positive correlation between $\Gamma$ and $E_c$ in Mrk~335, although we note that both the photon indices and cutoff energies ($<50$ keV) found there are small, differing substantially from other studies using the same datasets \citep{2014MNRAS.443.1723P,2015MNRAS.454.4440W}. The result we found here suggests a strong positive correlation between the photon index $\Gamma$ similar to Mrk~335.

\subsubsection{$Luminosity-E_c$ Correlation}\label{sec:l_ec}
Variability of the cutoff energy (or electron temperature) with flux or luminosity has not been explored in detail extensively in AGN, unlike black hole binaries. 
As we pointed out in section \ref{sec:gam_ec}, results from AGN are not conclusive yet, where both a correlation and an anti-correlation of $E_c$ and flux have been reported for 3C~382 and Mrk~335 respectively \citep{2014ApJ...794...62B,2016MNRAS.456.2722K}.

For X-ray binaries,
Cyg X-1 showed an increase in cutoff energy when the luminosity in the hard X-rays drops during the hard state \citep{1997MNRAS.288..958G,2005MNRAS.362.1435I}. A strong anti-correlation of the cutoff energy and luminosity was also observed in GX~339-4 when the luminosity was above $\sim10\%$ the Eddington luminosity, and it remained constant below that \citep{2008PASJ...60..637M}. A similar result was found by \cite{2009MNRAS.400.1603M}, who additionally observed the reverse trend when the source softened before transiting to the soft state. Similar behavior is seen in other objects, including V404~Cyg in its recent outburst as observed with \emph{Fermi} \citep{2016arXiv160100911J}, and possibly also Cyg X-1 in the soft state from \nustar observations \citep{2016arXiv160503966W}. We note that when $E_c$ is correlated with luminosity, so is $\Gamma$, and when the spectra soften during the hard intermediate state, \emph{both} $\Gamma$ and $E_c$ reverse their dependence on luminosity. The result we find for \mcg appears to match the behavior of black hole binaries not in the hard state where there $E_c$ is anti-correlated with $L$, but during the intermediate state. A correlation between $\Gamma$ and the flux is well established in bright AGN and Galactic black holes \citep[e.g.][]{2009MNRAS.399.1597S,2015MNRAS.447.1692Y}.

\subsubsection{Physical Interpretation}\label{sec:phys}
Before discussing the details of the plasma physics, it is worth mentioning the possibility that the changes in $E_c$ may not be due to changes in the intrinsic election temperature, but rather due to changes in the gravitational redshift of a constant spectrum \citep[e.g.][]{2016ApJ...821L...1N}. Such changes in the gravitational redshift of the emitted photons (due to changes in the size of the corona for instance) could artificially introduce variations in $E_c$ without the plasma properties changing. This, however, also produces changes in the reflection fraction due to the focusing of light rays into and away from the disk. The fact that the observed reflection fraction is constant suggests that the geometry does not change significantly, strengthening the interpretation in which the $E_c$ variations are \emph{intrinsic} to the plasma. Also, the inner radius we measure is relatively large, so GR effects are present but not extreme, and therefore the discussion of relativistic modeling in \citealt{2016ApJ...821L...1N} are not applicable in this work.

In the simplest considerations of a pure thermal plasma \citep{1980A&A....86..121S} (no electron-positron pairs), an increase in the soft flux impinging on the corona leads to softer Comptonization spectra and lower temperatures as the electrons are cooled efficiently \citep[e.g.][]{2002ApJ...578..357Z}. This picture cannot be applied directly here, first because pairs are not included and their effect could be important \citep{2015MNRAS.451.4375F}, and second because the correlation we measure is in the hard flux (emitted by the corona) rather than the soft flux impinging on it. Therefore, we would like first to assess the importance of pair production given the measurements we have.

We start by comparing the observations to predictions of pair-dominated plasmas. The temperature of a plasma cannot be arbitrarily high for a given size. The key parameter that is often used is the compactness $l=4\pi (m_p/m_e)(r_g/r)(L/L_{\rm edd})$, which measures the luminosity to source size ratio. As the compactness increases, photon-photon interactions become important, and any extra heating goes into producing electron-positron pairs rather than heating the plasma, causing the temperature to saturate \citep{1983MNRAS.205..593G,1985ApJ...289..514Z}. 

\begin{figure*}
\centering
\includegraphics[width=200pt,clip ]{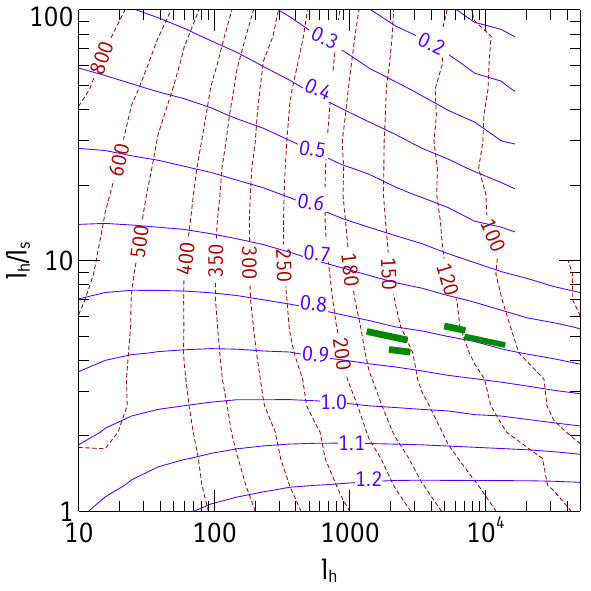} \hspace{1cm}
\includegraphics[width=200pt,clip ]{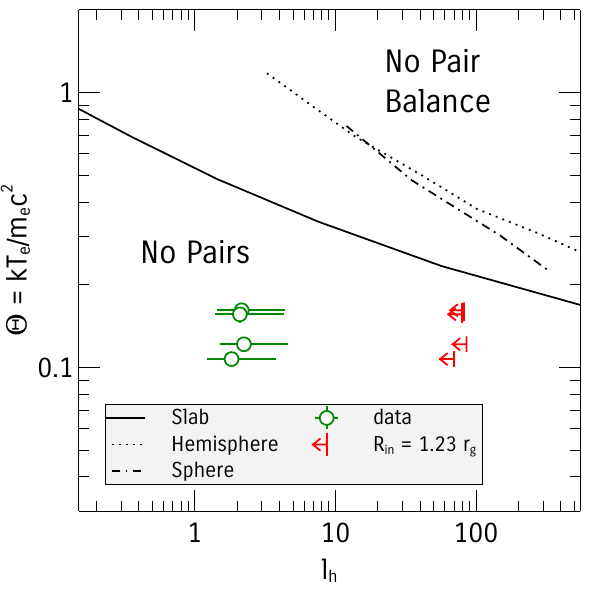} 
\caption{\emph{Left:} Contours of constant spectral index $\alpha=\Gamma-1$ (solid lines) and cutoff energy $E_c$ (in keV; dashed lines) for different parameters of compactness $l_h$ and hard to soft compactness ratio ($l_h/l_s$). The green boxes are the $1\sigma$ measurement error in $\alpha$ and $E_c$ from the four observations. This plot is essentially a conversion plot from the observed $\Gamma$ (or $\alpha$) and $E_c$ to the plasma parameters $l_h$ and $l_h/l_s$ using the \texttt{eqpair} model and assuming a plasma in pair equilibrium. \emph{Right:} The maximum temperature (in units of $m_e c^2$) that can reached by a plasma dominated by runaway pair production for three geometries. Upper limits to $l_h$ and measurements of $\Theta$ are shown.}
\label{fig:phys}
\end{figure*}

Following \cite{1985ApJ...289..514Z} and \cite{1994ApJ...429L..53G}, we calculate the spectral index $\alpha$ ($\alpha=\Gamma-1$) and cutoff energy predicted from models in thermal and pair equilibria, for different values of the compactness $l_h$ and $l_h/l_s$, where $l_h$ is the compactness of the Comptonization plasma (i.e. the power heating the corona), and $l_s$ is the compactness of the soft source producing the seed photons. We use the model \texttt{eqpair} \citep{1999ASPC..161..375C} to generate spectra for a grid of parameters for $l_h$ (between 10 and $5\times10^4$) and $l_h/l_s$ (between 1 and 100). We assume the soft source is a blackbody with a temperature of 10 eV and the plasma is spherical and contains no background electron plasma ($\tau_p=0$). The generated spectra are then fitted with a cutoff powerlaw model to simulate the fitting procedure and obtain the energy spectral index $\alpha$ and $E_c$. The results are shown in the left panel of Figure \ref{fig:phys}. Contours of $\alpha$ and $E_c$ are shown, and the $1\sigma$ measured values of $\alpha$ and $E_c$ are shown with the green boxes. This plot shows that, first, the inferred compactness ratio $l_h/l_s\sim5-6$, which is not atypical of AGN. Secondly and most importantly, it shows that, if the plasma is dominated by pairs, then the inferred $l_h$ is large (for reference, a source radiating at the Eddington limit that is 1 gravitational radius in size has $l\sim10^5$). Therefore, based just on the the measured $\alpha$ and $E_c$, if the plasma is dominated by pairs, the source has to be very compact suggesting a small size and/or high luminosity.

Further information is provided by including the source size measurement we have from the reflection spectrum and the observed luminosity. The results in this case are shown in the right panel of Figure \ref{fig:phys}, showing the temperature-$l$ relations at the pair limit from the modeling of \cite{1995ApJ...449L..13S} for three geometries of the corona. For a given compactness and a geometry, a source cannot have a temperature above the lines shown. Below the lines, the effect of pairs decreases and the plasma contains only electrons. As we have discussed in section \ref{sec:spec_mod}, the reflection spectrum constrains the inner radius of the disk to be $R_{\rm in}=47\pm23\, r_g$. If we assume the corona is of the same size (otherwise, relativistic features in the reflection will be washed out), we obtain the green circles shown in Figure \ref{fig:phys}. We used the cutoff powerlaw flux in the range 0.1--200 keV to measure the luminosity assuming a standard cosmology ($\Omega_m=0.3, \Lambda=0.7$) with $H_{0}=68\,km\, s^{-1}\, Mpc^{-1}$, and a black hole mass of $M=10^{7.9}\, M_{\odot}$ \citep{2012A&A...542A..83P}\footnote{We follow a similar procedure to \cite{2015MNRAS.451.4375F}.}. The red arrows show the upper limits on the observed $l_h$ obtained by setting $R_{\rm in}=1.23r_g$, the innermost stable circular orbit for a maximally spinning black hole. The electron temperature is estimated as $kT_e\sim E_c/2$ \citep{2001ApJ...556..716P}.

We can see that all the measurements fall below the pair limit lines for the three geometries, indicating that the plasma in \mcg is \emph{not} dominated by pairs and consists mostly of electrons. This is a robust statement given the small uncertainties in the cutoff measurements. Using the \texttt{nthcomp} \citep{1996MNRAS.283..193Z} model (as implemented in the reflection model \texttt{relxillCp} which calculates the reflection spectrum when illuminated by \texttt{nthcomp}; Garcia et al. \emph{in prep.}) instead of the \texttt{cutoffpl} shifts the electron temperatures up only by $\sim10\%$, and our conclusion about the plasma content is not altered. The points can of course be shifted to the right (i.e increasing $l_h$) if the black hole mass is erroneous. We find that in order for the plasma to be in pair balance for the slab geometry, the black hole mass needs to be smaller by $\sim$ two orders of magnitude (i.e. $M\sim10^6\, M_{\odot}$). We note that the black hole mass in \cite{2012A&A...542A..83P} is uncertain. A mass estimate using the fundamental plane of black hole activity \citep{2009ApJ...706..404G} using the radio fluxes from \citep{2009ApJ...703..802M} and the X-ray fluxes from our observations suggest a lower mass of $10^7\, M_{\odot}$. Other estimates suggest similar smaller values \citep[e.g.][]{2010ApJ...720L.206Z}, but not low enough to alter our conclusions about the plasma content. Note also that the correlation between $E_c$ and $l_h$ is weaker than the $E_c-F_p$ correlation in Fig. \ref{fig:ecut_var}. That is because we use the wider energy range to calculated the flux and also because of the uncertainties in the radius measurements.

\subsubsection{Cutoff Variability}
The additional information provided by the four measurements of the plasma properties provide further constraints. The cutoff energy in pair-dominated plasmas scales inversely with compactness (or with luminosity when the source size is constant, as is the case here where we measure a constant reflection fraction and reflection parameters): $E_c\propto l_{h}^{-1}$ \citep{1994ApJS...92..585S}. This comes from the fact that increasing $l_h$ produces more pairs, and for balance to hold, the same energy is now distributed to more particles so the energy per particle drops. This trend is \emph{not} what we observe in \mcg. Instead we find that the cutoff energy is \emph{higher} for higher fluxes, providing further evidence that the plasma is not dominated by pairs. In electron plasmas on the other hand, the electron temperature is not expected to depend on $l_h$ for a given ratio $l_h/l_s$ and optical depth $\tau$. This comes from the fact that although electrons gain energy when $l_h$ increases, the constant $l_h/l_s$ means cooling also increases so to keep the temperature constant. The fact that the cutoff energy (or electron temperature) varies with on $l_h$ observationally means that either $l_h/l_s$ or $\tau$ is variable. The first is ruled out by virtue of Fig. \ref{fig:phys}-left. We conclude, therefore, that the optical depth $\tau$ varies in a way to produce the $E_c-$flux relation in Figure \ref{fig:ecut_var}.



\begin{figure}
\centering
\includegraphics[width=200pt,clip ]{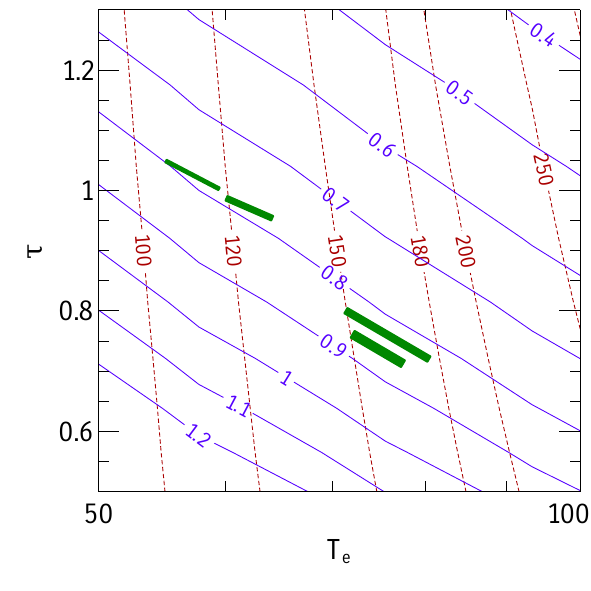}
\caption{Contours of constant spectral index $\alpha=\Gamma-1$ (solid lines) and cutoff energy $E_c$ (in keV; dashed lines) for different parameters of optical depth $\tau$ and electron temperature (in keV). The green boxes are the $1\sigma$ measurement error in $\alpha$ and $E_c$ from the four observations. This plot is essentially a conversion plot from the observed $\Gamma$ (or $\alpha$) and $E_c$ to the plasma parameters $\tau$ and $T_e$ using the \texttt{compps} model and assuming a slab geometry.}
\label{fig:phys_compps}
\end{figure}

We now turn to the $E_c-\Gamma$ correlation.
In pair-dominated plasma, and for temperatures below the rest energy of the electron (as observed here), $E_c$ is expected to be either positively correlated or independent of $\Gamma$, depending on the temperature and the compactness $l_h$ \citep{1994ApJ...429L..53G,2002ApJ...578..357Z}. In pair-free plasma on the other hand, an $E_c-\Gamma$ anti-correlation is expected for a fixed $l_h$. However, as we have already established observationally, $l_h$ is not constant, which implied a variable optical depth $\tau$. To investigate the variable $\tau$ possibility further in pair-free plasma, 
we employ a similar method to that used to produce Figure \ref{fig:phys} and discussed in \ref{sec:phys}, but now we use the model \texttt{compps} \citep{1996ApJ...470..249P}, and use a grid of $T_e$ and $\tau$. The results are shown in Figure \ref{fig:phys_compps} for a slab geometry (geometry parameter in \texttt{compps} set to 1). We find that the optical depth $\tau$ varies significantly with $T_e$, being lower for higher temperature. The exact values of $\tau$ depend on the assumed geometry (e.g. $\tau$ varies between 1.3--1.9 if we assumed a spherical geometry instead of the slab geometry shown in Figure \ref{fig:phys_compps}).

The conclusion here is that, in order to explain the $E_c-\Gamma$ and $E_c-F_p$ correlations with, the plasma need to be pair-free and its optical depth need to change as shown in Figure \ref{fig:phys_compps}. This could possibly be accompanied by geometry changes, but it has to be small enough for the inner radius and $R$ measurements to remain constant within the observational uncertainty.

One additional observation that can be noted from Fig. \ref{fig:phys}-left, is that the ratio of heating to cooling in the corona changes very little between observations. The observed changing coronal flux therefore suggests that the photon flux cooling the corona \emph{changes} too, and in the same direction.
This could be achieved if the UV photons of the disk vary with the X-rays. The UV flux from the source measured with the \swift UVOT camera however, shows little variations compared to the X-rays (the fractional RMS variation in X-rays is $24\pm2\%$ while in the UV, it is $10\pm4\%$ in the W2 filter and no more than $5\%$ in other bands red-ward of $2000{\rm\AA}$). The soft flux reaching the corona could in principle change if only the disk temperature changes, so the flux in the UVOT filters is not affected. This however would suggest that the inner radius changes, which not observed. The constant heating over cooling we find implies that there is a feedback between the hot corona and the disk photons cooling it \citep[e.g.][]{1991ApJ...380L..51H}, suggesting that a significant part of the photons cooling the corona are due to reprocessing in the disk.

Comparing our results with \cite{2015MNRAS.451.4375F} indicates that AGN coronae are not always pair-dominated, or that some sources are in that regime and others are not. Mrk~335 appears to show a behavior similar to that reported here for \mcg \citep{2016MNRAS.456.2722K}, so it too, is unlikely to be pair-dominated. The presence or absence of reflection close to the black hole is unlikely to be be the reason of the difference, nor is the Eddington ratio (Mrk~335 is accreting close to the Eddington limit while \mcg accretes at the few percent level). Studies of other sources with \nustar in the near future will help address the issue. 

\section{Conclusion}
We use data from the longest \nustar observing campaign of a Seyfert galaxy to study the properties of the plasma responsible for the hard X-ray emission. The sensitivity of \nustar allows us to constrain the plasma properties and probe its variability. Our main results are as follows:
\begin{itemize}
\item The inner radius of the disk and its emissivity remain constant between observing epochs, suggesting a constant geometry. Most of the spectral variability is due to changes in the flux and spectral index of the primary X-ray source. Flux from the relativistic reflection follows the flux from the direct component.
\item The measured cutoff energies (and inferred electron temperatures) are not high enough for the plasma to be dominated by electron-positron pairs, unless the black hole mass is two orders of magnitude lower. This means that the plasma contains mostly electrons.
\item We find that the cutoff energy is strongly correlated with both the source flux and the spectral index. The former correlation is another indication that the plasma is not dominated by pairs. The two correlations are driven by changes in the optical depth of the plasma.

\item A constant heating to cooling ratio is inferred for the plasma. This, along with the constant UV flux observed,
suggest a feedback mechanism in which most of the photons cooling the hot corona are due to reprocessing in a cold disk.
\end{itemize}

\section*{Acknowledgment}
We thank the referee A. Zdziarski for the useful comments and suggestions
that helped with the interpretation of the data.
This work has been partly supported by NASA grant NNX14AF89G.
This work made use of data from the \nustar mission, a project
led by the California Institute of Technology, managed by the Jet
Propulsion Laboratory, and funded by the National Aeronautics and
Space Administration. We thank the \nustar Operations, Software
and Calibration teams for support with the execution and analysis of
these observations. This research has made use of the \nustar
Data Analysis Software (NuSTARDAS) jointly developed by the ASI
Science Data Center (ASDC, Italy) and the California Institute of
Technology (USA).

\bibliographystyle{astron}
\bibliography{main}

\end{document}